\documentclass[%
 aip,
 amsmath,amssymb,%nobibnotes,
preprint,%
]{revtex4-2}

\usepackage{graphicx}% Include figure files
\usepackage{dcolumn}% Align table columns on decimal point
\usepackage{bm}% bold math
\usepackage[utf8]{inputenc}
\usepackage[T1]{fontenc}
\usepackage{mathptmx} 
\usepackage{float}
\binoppenalty=\maxdimen
\relpenalty=\maxdimen

\begin{document}

\title{Engineering the reciprocal space for ultrathin GaAs solar cells}% Force line breaks with \\

\author{Jeronimo Buencuerpo}
\affiliation{National Renewable Energy Laboratory (NREL), 15013 Denver W Pkwy, Golden, CO 80401, USA}
\email{jeronimo.buencuerpo@nrel.gov}
\author{Jose M.~Llorens}
\affiliation{Instituto de Micro y Nanotecnología, IMN-CNM, CSIC (CEI UAM+CSIC) Isaac Newton, 8, E-28760, Tres Cantos, Madrid, Spain}
\author{Jose M.~Ripalda}
\affiliation{Instituto de Micro y Nanotecnología, IMN-CNM, CSIC (CEI UAM+CSIC) Isaac Newton, 8, E-28760, Tres Cantos, Madrid, Spain}
\author{Myles A.~Steiner}
\affiliation{National Renewable Energy Laboratory (NREL), 15013 Denver W Pkwy, Golden, CO 80401, USA}
\author{Adele C.~Tamboli}
\affiliation{National Renewable Energy Laboratory (NREL), 15013 Denver W Pkwy, Golden, CO 80401, USA}

\begin{abstract}
III-V solar cells dominate the high efficiency charts, but with significantly higher cost than other solar cells.
Ultrathin III-V solar cells can exhibit lower production costs and immunity to short carrier diffusion lengths caused by radiation damage, dislocations, or native defects. Nevertheless, solving the incomplete optical absorption of sub-micron layers presents a challenge for light-trapping structures. Simple photonic crystals have high diffractive efficiencies, which are excellent for narrow-band applications. Random  structures a broadband response instead but suffer from low diffraction efficiencies. Quasirandom (hyperuniform) structures lie in between providing high diffractive efficiency over a target wavelength range, broader than simple photonic crystals, but narrower than a random structure. 
In this work, we present a design method to evolve a simple photonic crystal into a quasirandom structure by modifying the spatial-Fourier space in a controlled manner. We apply these structures to an ultrathin GaAs solar cell of only 100 nm. We predict a photocurrent for the tested quasirandom structure of 25.3 mA/cm$^2$, while a planar structure would be limited to 16.1 mA/cm$^2$. 
The modified spatial-Fourier space in the quasirandom structure increases the amount of resonances, with a progression from discrete number of peaks to a continuum in the absorption. The enhancement in photocurrent is stable under angle variations because of this continuum. We also explore the robustness against changes in the real-space distribution of the quasirandom structures using different numerical seeds, simulating variations in a self-assembly method. 
%We predict that the reciprocal space results in a mean  photocurrent of 25.0 mA/cm$
%^2$. 
We observe a standard deviation of the quasirandom structures of only 0.3 mA/cm$^2$.
The approach presented here can be applied to guide future optimizations and experimentalists to design and fabricate new classes of photonic crystals for GaAs and other ultrathin solar cells.
%The standard deviation of the photocurrent is only of 0.2 mA/cm$^2$. This deviation is comparable to real space defined structures when allowing for fabrication errors in top-down approaches.
\end{abstract}

\maketitle 
\section{Introduction}

Ultrathin III-V cells have the potential to reduce deposition time and material usage while keeping high efficiency. Both are fundamental to lower the cost.\cite{horowitz__2018}
They also allow for higher efficiencies when carrier diffusion lengths are a limiting factor, such as in dilute nitrides,\cite{ochoa_AIPConferenceProceedings_2016} radiation damaged space solar cells,\cite{maximenko_2019IEEE46thPhotovolt.Spec.Conf.PVSC_2019,mellor_Sol.EnergyMater.Sol.Cells_2017,hirst_Appl.Phys.Lett._2016,yamaguchi_SolarEnergyMaterialsandSolarCells_2001} 
 or metamorphic materials with high dislocation densities\cite{mehrotra_2014IEEE40thPhotovolt.Spec.Conf.PVSC_2014,wilson_AIPConferenceProceedings_2016}.
However, to achieve competitive efficiencies, ultrathin devices require optical engineering, namely, light trapping. Our objective is to reduce the cell thickness without sacrificing optical absorption.

The typical light-trapping structures are either ordered 2D lattice patterns
like photonic crystals (PC), or completely disordered
patterns.\cite{brongersma_Nat.Mater._2014} Simple periodic structures are highly
diffractive because of the limited number of diffraction orders. They are
excellent couplers for narrow band applications. In the critical coupling regime
it is possible to reach a 100\% absorption, but they fall below the light
trapping limit for broadband applications like
photovoltaics.\cite{yu_Proc.Natl.Acad.Sci._2010}
In contrast to simple PC, completely random patterns, like self-assembly roughness, are weakly diffractive but have broadband spectral features (they can diffract light in a wavelength range).\cite{ferry_NanoLett._2011} In between these two limits are the quasirandom (QR), or hyperuniform photonic crystals.\cite{yu_Sci.Rep._2017,martins_NatCommun_2013,castro-lopez_APLPhotonics_2017,gorsky_APLPhotonics_2019} They have a richer Fourier spectrum than a simple photonic crystal and a higher diffraction efficiency than a complete random structure. This feature makes them more suitable for broadband applications. 

In this work, we present a general approach for designing photonic crystals valid
not only for simple periodic structures but also for quasirandom structures.
This transition is done by increasing the density of propagating Bragg harmonics
in the reciprocal space, namely the power spectral density (PSD). There are
different approaches to study the evolution between an ordered structure, like
a simple PC, to a random structure.\cite{ferry_NanoLett._2011,vynck_NatMater_2012} A simple approach is to disorganize (introducing random displacements to the initial periodic positions) multiple unit cells of the PC, creating a super-cell. \cite{bozzola_Prog.Photovolt:Res.Appl._2014,oskooi_ApplPhysLett_2012,vynck_NatMater_2012, ferry_NanoLett._2011,ding_Opt.Express_2016} A refinement of this approach is to randomize the super-cell while aiming at a target PSD distribution. \cite{martins_NatCommun_2013,vanlare_ACSPhotonics_2015,gorsky_APLPhotonics_2019,xiao_Opt.Express_2018} These techniques heavily rely on the real space distribution, and the
optimization algorithm used. Furthermore, these methods  are computationally intensive, limiting the study for big unit cells. 

The alternative to real space design is to work directly in the reciprocal space and, later, obtain the final unit cell in the real space.  The ordering imposed by  design defines the optical response of the structure.\cite{yu_Sci.Rep._2017,ma_JournalofAppliedPhysics_2017} 

The reciprocal technique used in Refs.\citenum{yu_Sci.Rep._2017,ma_JournalofAppliedPhysics_2017} is the Gaussian random field (GRF). This technique is fast and computationally efficient when creating big unit cells. Yet, it does not allow a transition from a simple PC to a QR. This limitation complicates the analysis of the transition from a sparse PSD typical of a simple PC, to a denser PSD as found in QR. 

In this paper, we propose an alternative design method based on the iterative Fourier transform algorithm (IFTA) \cite{gerchberg_Optik_1972, ripoll_OE_2004}. 
The IFTA methods are broadly adopted in holography as they provide an effective way to design 2D diffractive optical elements (DOE). It has been explored also for random textured surfaces following the scalar theory,\cite{rockstuhl_Opt.ExpressOE_2010} but not developed for binary photonic crystals, neither periodic or quasi-random.
There is an analogy between a DOE and its hologram, with a light trapping structure and its PSD; both are related by the Fourier transform. Therefore, using IFTA we can design in-plane nanostructures from a controlled PSD in reciprocal space. 

In Section~\ref{sec:design} we provide the details of a 100 nm GaAs solar cell used to illustrate the absorption enhancement of our proposal. The nanostructure is outside of the solar cell active area to avoid parasitic surface recombination losses. We introduce the rationale behind the target PSD and describe the IFTA implementation. We show that it is possible to modify the sparsity of the PSD generating structures from simple PC to QR with our approach. The results are presented in Section~\ref{sec:results}. We analyze the absorption and asses the performance of the solar cell calculating the short-circuit current. We identify the advantages of each PSD provides, their best current and explore the robustness against changes in the real space distribution of the QR structures. We find the enhancement from the  QR structure is stable under angle variations.%We find that the QR structure obtains higher photocurrent than their simple PC counterparts.

\section{Photonic Design}
\label{sec:design}

\subsection{Solar cell structure}
\begin{figure*}[htbp]
\centering
    \includegraphics[scale=1.]{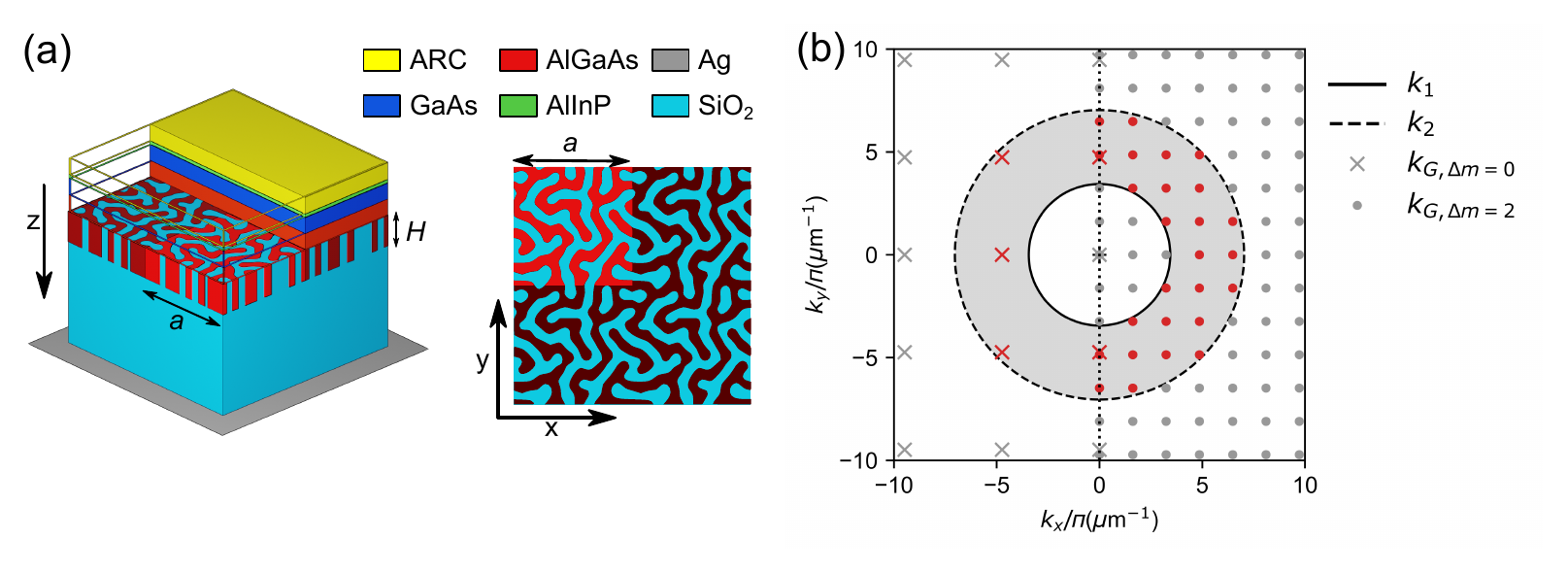}
\caption{(a) Scheme of the layer structure of the solar cell, with a conventional ARC (yellow, 100 nm MgF$_2$ and 50 nm TiO$_2$) a 10 nm window layer of AlInP (green) an ultrathin GaAs absorber  of 100 nm (blue),  a back surface field layer of AlGaAs (red), and back nanostructure of AlGaAs (red) in SiO$_2$ (blue) over a silver mirror (gray). The nanostructure is placed in the $x$-$y$ plane, and with periodic boundary conditions; light is incident along the z direction (b) Outer limit of Bragg harmonics, for frequencies associated with the wavelengths of $\lambda_g=900$ nm, $k_1$, (black solid), $\lambda_\mathrm{sp}=440$ nm, $k_2$, (black dash-dotted) inside the SiO$_2$. The gray filled area  between $k_1$ and $k_2$ is the design objective. The grey crosses are the Bragg harmonics for a lattice of 422 nm ($\Delta m=0$, for clarity only plotted for negative $k_{G,k_x}$), and the grey dots are the Bragg harmonics for a lattice of 1230 nm ($\Delta m=2$, for clarity only plotted for positive $k_{G,k_x}$). The red dots and crosses are the target spectral frequencies.}
\label{fig:design}
\end{figure*}

The  solar cell layer structure considered in this study is shown in Fig.~\ref{fig:design}. The photocurrent ($J_\mathrm{sc}$) is generated in a 100 nm GaAs layer. Such cell is in the ultrathin regime
in the sense of incomplete absorption without photonic aid and sub-wavelength thickness. 
It is of particular interest because this thickness allows effective radiation-hard devices.\cite{hirst_Appl.Phys.Lett._2016,maximenko_2019IEEE46thPhotovolt.Spec.Conf.PVSC_2019}
We use a standard double layer antireflective coating (ARC) placed on top of the absorbing layer. The ARC is a 100 nm MgF$_2$ layer on top of a 50 nm TiO$_2$ layer, both layers shown as a yellow slab in Fig.~\ref{fig:design}(a). The cells is modeled as a 10 nm AlInP window layer, a 100 nm GaAs homojunction (active region) and a 100 nm back surface field layer, BSF, of Al$_{0.8}$Ga$_{0.2}$As. We place the PC of square lattice constant $a$, and height $H$ just beneath the back surface field. Choosing the material comprising the PC is key to achieve a high $J_\mathrm{sc}$. The real part of the refractive index ($n$) mostly impacts on the scattering efficiency. Meanwhile, the imaginary part ($\kappa$) leads to undesired parasitic absorption, as the absorption in the PC does not contribute to the photocurrent. The high refractive index component of the PC is Al$_x$Ga$_{(1-x)}$As, and the low index component is SiO$_2$. The AlGaAs systems offer tunability of the optical properties by controlling the aluminum content and a high $n$.\cite{buencuerpo_Opt.ExpressOE_2020} We use Al$_{0.8}$Ga$_{0.2}$As, as a compromise to minimize $\kappa$ and maximize $n$. 
Placing the nanostructure in the back of the cell results in a design more forgiving for parasitic absorption. A large fraction of the high energy photons are absorbed before reaching the PC allowing us to use high index semiconductors instead of transparent dielectrics. Dielectrics present lower absorption than the semiconductors but also a lower refractive index. Finally, the structure is terminated by a 1$\mu$m SiO$_2$ spacer and a silver mirror. The dielectric spacer is key to mitigating the silver's parasitic absorption and leads to two-pass diffraction in each reflection, increasing absorption in the thin GaAs layer. 
The current extraction in this design is assume to be done using hundred of microns to millimeters spaced fingers, or point contacts, in the front and back of the solar cell,\cite{chen_NatEnergy_2019,eerden_Prog.Photovolt.Res.Appl._2020}. This approach is posible thanks to the good lateral conductivity of III-V materials in contrast with other thin-film materials. The shadowing from these contacts is not take in account, typically representes a 2-3\% current loss just for geometrical reasons, and it applies for all the structures. Nevertheless, this losses can be minimized by ray-optics,\cite{saive_AdvancedOpticalMaterials_2016} or subwavelength structures\cite{san_roman_cloaking_2016}. 

We use rigorous coupled wave analysis (RCWA),\cite{liu_Comput.Phys.Commun._2012} for modeling the optical response of the structures, namely the reflection and absorption of each layer. We also do a further analysis using the Fourier series that reconstructs the fields in the unit cell to understand the contributions to the absorptance in the GaAs layer ($A_{\mathrm{GaAs}}$) and to the reflection ($R$).\cite{buencuerpo_Opt.ExpressOE_2020} 
\begin{align}
A_{\mathrm{GaAs}} = A_{\mathrm{GaAs},g=0} + A_{\mathrm{GaAs},g>0},\\
R = R_{g=0} + R_{g>0},
\end{align}
where $g=0$ denotes the contribution from the non-scattered light and $g>0$ that of the scattered light. In other words, we split the light that propagates normal to the solar cell's surface and that of light that changes its wavevector by the PC.  

We will analyze the predicted device performance in the radiative limit using the generated photocurrent, $J_\mathrm{sc}$,\cite{nelson__2003} identifying the $A_\mathrm{GaAs}$ with the cell's external quantum efficiency. Analogously, we will define the scattered contribution to the $J_{\mathrm{sc},g>0}$ using only  $A_{\mathrm{sc},g>0}$. We have taken the refractive index from Ref. \cite{palik__1991,tanemura_AppliedSurfaceScience_2003,aspnes_JournalofAppliedPhysics_1986}. 

\subsection{Target power spectral density}
The main design task is to find the real space PC whose power spectral density
(PSD) in the reciprocal space is negligible everywhere except in the region of interest. The fundamental idea is that by enhancing the PSD of only a set of meaningful wavevectors $\bm{k}$ we can diffract the propagation of the light inside the solar cell and therefore increase the optical path. Fig.~\ref{fig:design}(b) schematically shows the allowed wavevectors of a square lattice PC. The area between the two circles encloses the associated diffraction orders we want to excite. Conversely, we want to minimize the diffraction associated to those outside. It is important to excite counter propagating orders to avoid polarization sensitivity issues, hence the circle geometry. The hollow disk is defined by two wavevectors constraining the spectral absorption range. The shortest wavevector is determined by the bandgap of the GaAs, $870$ nm. We will round up to $\lambda_g = 900$ nm to include the Urbach-tail absorption of the GaAs. The largest one corresponds to $\lambda_\mathrm{dp}= 440$ nm, as full absorption is obtained in a double pass through the 100 nm GaAs layer for $\lambda<\lambda_\mathrm{dp}$. 

From these two limiting wavelengths, we can proceed to adjust the wavevectors to the materials encapsulating the PC: AlGaAs and SiO$_2$, see Fig.~\ref{fig:design}(a).
Hence, we have to adjust both wavevectors using the refractive index. For simplicity, we choose a constant refractive index of n=1.4, value in the bottom limit for SiO$_2$, as the PC will diffract light transmitted into the SiO$_2$.   
Upon reflection at the silver back mirror, the reflected light will be diffracted again in the PC crystal before reentering the GaAs layer. This double diffraction increases the efficiency of our design. 
Therefore, the range of $k$ limits is defined by,
    $k_1 = k_{g,\text{SiO}_2}$, and 
    $k_2 = k_{\text{dp}, \mathrm{SiO}_2}$,
$k_1$ and $k_2$ are the lower and upper limits, respectively.

The number of reciprocal lattice vectors contained in the ring spanned between $k_1$ and $k_2$ is discrete and defined by the periodicity of the PC, $a$. The in-plane wavevector is $\bm{k}_{\bm{G}}=(k_{x, \bm{G}}, k_{y,\bm{G}})$. Where $k_{x, \bm{G}}=k_x + G_x$ and $k_{y,\bm{G}}=k_y + G_y$. Therefore, the reciprocal lattice vector is defined by the lattice constant $a$ as $\bm{G}=2\pi/a (g_x, g_y)$, where $g_{x,y}=0, \pm1, \pm2,\ldots\pm N$.
The minimum distance in the reciprocal lattice is $\Delta G = 2\pi/a$, see Fig.~\ref{fig:design}(b).  

The objective ring limits are set by $k_1$ and $k_2$, but not the number of reciprocal vectors inside it. We define $\Delta m$ as the number of discrete frequencies in the objective ring radius. $\Delta m > 0$ for $\Delta G=(k_2 - k_1)/\Delta m$, and
$\Delta m=0$ is obtained from $\Delta G=(k_1+k_2)/2$, see Fig. \ref{fig:design}(b). 
Bigger $\Delta m$ leads to bigger unit cells. 
Once we have defined $k_1$, $k_2$, $a$  and $\Delta m$ we can create the objective discrete reciprocal space, $Q$. 

\subsection{Iterative Fourier Transform Algorithm}
The Iterative Fourier Transform Algorithm (IFTA) \cite{gerchberg_Optik_1972,ripoll_OE_2004} can generate diffractive optical elements (DOE) from an objective intensity, and it is mainly used for  micrometric diffractive features and visible monochromatic beams. IFTA algorithms quickly converge compared with global based random optimization.\cite{birch_OpticsandLasersinEngineering_2000}. However, in our context, the DOE is  subwavelength, the PC,  and it introduces the diffraction at different wavelengths of light. Thus, the necessity of adapting the method and setting first the reciprocal space target to broadband applications.
 
The algorithm starts by defining the discrete wavectors Fourier space target, $Q$. $Q$ is a square matrix with $(2N+1)\times (2N+1)$ elements, which identifies with the PSD. Then, we normalize $Q$, with maximum values one and minimum values zero. Then, we add an imaginary component to the objective $Q$, $K^{0} = Q + iS$, where $S$ is a matrix of uniformly distributed random numbers. The seed for the first iteration can be initialized with a constant value or pre-condition distributions. Using a random seed improves the final convergence to the desired target $Q$. Once the algorithm is initalized, the iteration for each step, $s$, is defined as follows:
\begin{enumerate}
\item Back-propagate with the inverse discrete Fourier transform of $K^s$, from the reciprocal space into the matrix $U = \mathcal{F}^{-1}(K^s)$.
\item Compute the element-wise phase of $U$, $\Phi = \arg(U) = \text{atan2}(\Im(U)/\Re(U))$.
\item Quantize the phase matrix $\Phi$ on two levels (0 and $\pi$).
\item Propagate the phase (foward to the reciprocal space) using the Fourier transform into the matrix $D^s = \mathcal{F}(\exp[i\Phi])$ 
\item Test the difference,  $\delta$, between the objective image, $Q$, and the $s$ approximation: \\
$\delta= \sum_{i,j} ||Q_{i,j}| - |D^{s}_{i,j}||$. The exit condition is satisfied when the minimum tolerance for $\delta$ is met or the maximum number of steps is reached. 
\item Otherwise, the next iteration $K^{s+1} = Q \circ \exp(i \arg[D^{s}])$ is the element-wise product of the objective distribution and the phase.  
\end{enumerate}
Our structure is the final binary phase matrix $\Phi$ with two levels. We will associate the zero phase to SiO$_2$ and the $\pi$ phase to  AlGaAs, see Fig.~\ref{fig:design}(a).  

IFTA stagnates quickly, namely $\delta$ do not meet the exit criteria, especially for low-resolution images.\cite{ripoll_OE_2004, wyrowski_J.Opt.Soc.Am.AJOSAA_1988} 
A simple solution is  to expand the objective image, $Q$ in our case, creating a frame filled with zeros for higher frequencies than the target $Q$, namely a zero-padding applied before the start of the algorithm.\cite{ripoll_OE_2004} The zero-padding increases the level of detail in real space by adding a frame of higher spatial frequencies.
$\delta$ is computed on a constrained set of relevant $\bm{k}_{\bm{G}}$ vectors, namely leaving this frame outside of the convergence criteria. Then, the algorithm runs as described. 
Nevertheless, the zero-padding technique might end up in a fragmented structure, as we have introduced high spatial frequency components, which complicates a later fabrication. We applied IFTA twice, with and without zero-padding, using the first result as the seed in the second iteration. We find this approach is stable against stagnation and fragmentation. 

\section{Results} 
\label{sec:results}
\subsection{Unit cells created}
\begin{figure*}[htbp]
\centering
    \includegraphics[scale=0.8]{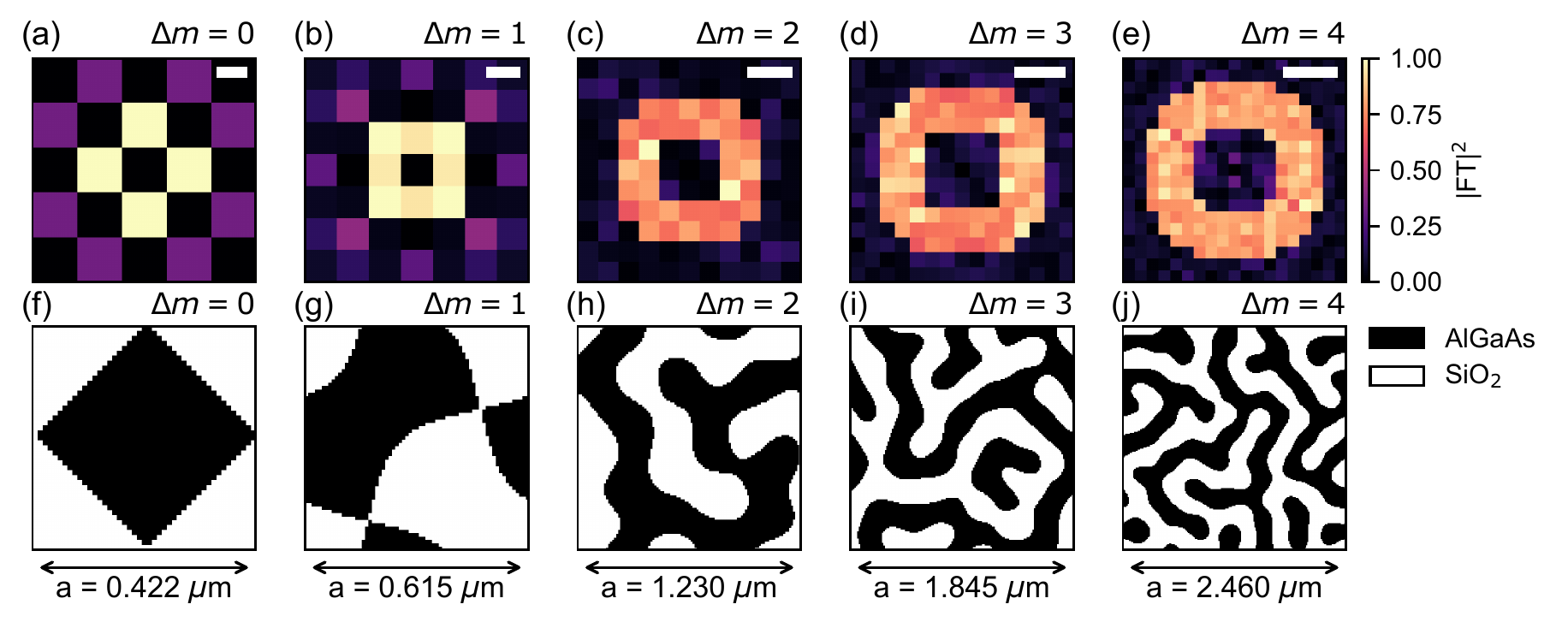}
\caption{(a)-(e) Reciprocal space for orders different than zero for an increasing number of $\Delta m$. The inset bar is 4$\pi$ $\mu$m$^{-1}$. (f)-(j) Real space reconstruction of the $k$-space designs on top. It represents the unit cell  of the light trapping layer made of SiO$_2$ (white) and AlGaAs (black).}
\label{fig:kgrowth}
\end{figure*}

Once we have found a systematic way of generating our target $Q$,  we start by exploring the impact of $\Delta m$ on the performance of a solar cell. We have only considered values $\Delta m\le4$. The generated structures are shown in Fig. \ref{fig:kgrowth}. 
Increasing $\Delta m$ results in a better defined ring of increasing width, see Figs.~\ref{fig:kgrowth}(a-e). The rendering in real space is shown in Figs.~\ref{fig:kgrowth}(f-j). The first thing to note is the increase in the lattice constant, hence significantly larger structures are needed to obtain a larger number of diffraction orders. Also, the patterns experience an evolution. When using $\Delta m=0$ we obtain a typical chess-board in square lattice, see Fig.~\ref{fig:kgrowth}(f). We create the thinnest possible ring for $\Delta m=1$. It presents a Fourier spectrum similar to a conventional square lattice but exciting the diagonal first orders to form the ring. The condition of $\Delta m=1$ generates an interesting PC with bow-tie appearance. 
The successive cases with $\Delta m > 1$ produce the deterministic QR structure already mentioned in the introduction.\cite{martins_NatCommun_2013,vanlare_ACSPhotonics_2015,yu_Sci.Rep._2017} 

The modulus of the reciprocal space, the orders $(g_x, g_y)$, will define the structure's anisotropy, as they represent the diffraction orders being excited by the PC. The structures of $\Delta m =0$ present a PSD symmetric with respect to $k_x$ and $k_y$ axis. These structures \emph{always} exhibit a square lattice symmetry.
The QR structures are not entirely isotropic, as there are small differences in the PSD of the $(g_x, g_y)$ orders inside the ring. 
However, these differences appear to be random, and therefore the fluctuation compensate on average, after considering the contribution of all orders in the RCWA calculation.
In any case, when analyzing the optical response, we analyze $s$ and $p$ polarizations, and add them equally for taking into account any present anisotropy.  

\subsection{Absorption and photocurrent analysis}
\begin{figure*}[t] 
    \centering
   \includegraphics[scale=1]{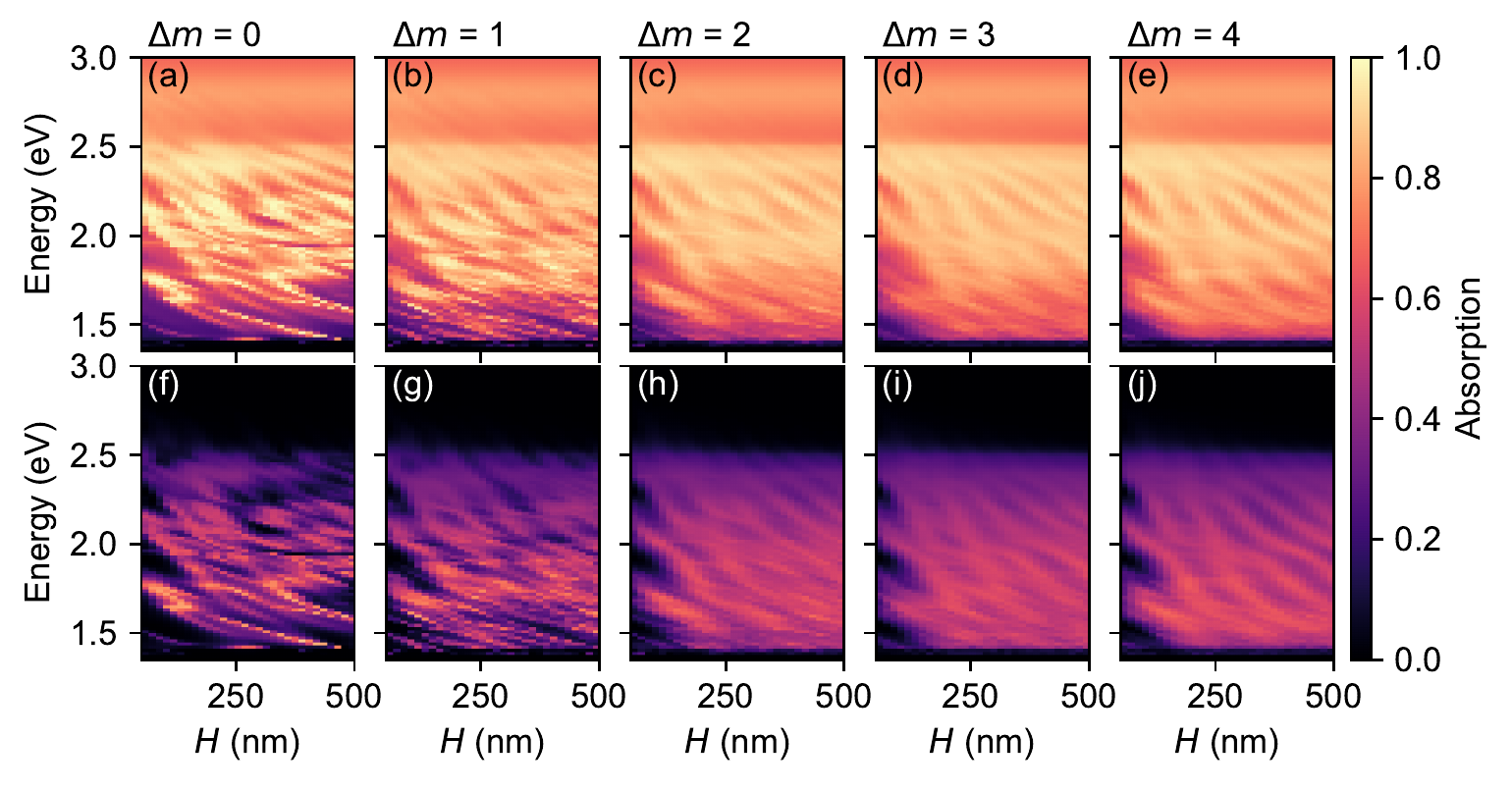}
\caption{First row, (a)-(f) Absorption in the GaAs layer, $A_\mathrm{GaAs}$, for structures with $\Delta m$ 0 to 4, respectively. Second row, (g)-(l), scattered contribution to the absorption in the GaAs layer, $A_{\mathrm{GaAs},g>0}$, for the same structures as the first row.}
\label{fig:absmap}
\end{figure*}

The thickness of the layer containing the diffractive element is not given by the design procedure described in Section~\ref{sec:design}. However, the optical response of the stack has a fundamental dependence on the structure thickness $H$. To show this dependence, we study the absorption in the GaAs layer sweeping the thicknesses of the PC from 50 nm to 500 nm (30 points). 
We show the corresponding spectra in Fig.~\ref{fig:absmap} as contour maps for energies up to 2.5 eV to put the focus on the resonant low energy absorption peaks, but we use up to 3.5 eV to calculate the $J_{\mathrm{sc}}$. 

\begin{table}[htbp]
  \centering
  \caption{Scale factor $N$ ($Q$ rank $2N+1$), Lattice parameter $a$, and $J_\mathrm{sc}$ for AM1.5g obtained for increasing $\Delta m$ photonic crystals using a GaAs absorber of 100 nm. For comparison, the predicted photocurrent for a reference structure without PC an ultrathin, 100 nm, is 16.1 mA/cm$^2$.  $H_\text{opt}$ is the optimal thickness from Fig.\ref{fig:jsc_thickness} analysis. The losses in the window layer absorbs $L_w=$ 0.7 mA/cm$^2$ for all the cases studied. }
\addtolength{\tabcolsep}{-1pt}   
\begin{tabular}{lccccr}
    $\Delta m$ & 0     & 1     & 2     & 3     & 4    \\
    \hline
    N & 2 & 3 & 5 & 7 & 9 \\
    $a$ (nm)     & 422  & 615 & 1230  & 1845  & 2460  \\
    $H_\mathrm{opt}$ (nm) & 376 &  252 & 453 & 236 & 205  \\
    $J_\mathrm{sc}$(mA/cm$^2$)  & 23.2  & 24.2  & 24.9  & 24.9  & 25.3 \\  
    $J_{\mathrm{sc},g>0}/J_\mathrm{sc}$ & 0.43 & 0.47  & 0.53  & 0.53  & 0.54 \\ 
    $L_\mathrm{AlGaAs}$ (mA/cm$^2$) & 0.6 & 0.6 & 0.7 & 0.6 & 0.5 \\
    $L_\mathrm{Ag}$ (mA/cm$^2$)& 0.2 & 0.2 & 0.3 & 0.2 & 0.3\\
    \hline
    \end{tabular}%
\addtolength{\tabcolsep}{1pt}   
  \label{tab:lattice_jsc}%
\end{table}%

The five nanostructures present the same valleys in $A_{\mathrm{GaAs},g>0}$ for shallow structures, i.e. thickness below 100 nm. These valleys are the resonances of the zero diffraction order (ZDO), namely the Fabry-Perot (FP) resonances. 
In other words, when a thin nanostructure is not able to diffract or scatter light, all the absorbed light comes from the ZDO. In between these valleys, the $A_{\mathrm{GaAs},g>0}$ increases due to the diffraction orders that get opened in the optical stack. The ZDO resonances are common to the five structures. Despite the different geometries, all the structures present the same materials and the same filling factor of the unit cell (around 50\%). The ZDO treats the photonic crystal as a medium of effective refractive index, and, therefore, all the nanostructures exhibit the same resonance features. However, in structures of higher thicknesses, there is a different distribution of peaks in each case. Indeed, the differences are more pronounced as the valleys from the FP resonances get blurred for $H>$  100 nm.

The absorption of the chess-board square structure, $\Delta m=0$, (Figs. \ref{fig:absmap}(a) and (f)) presents peaks of high intensity but of narrow bandwidth. The absorption peaks are even more evident when splitting the absorption contributions as the FP resonances are decoupled when studying only the scattered absorption, $A_{\mathrm{GaAs},g>0}$. These peaks are intense but narrow, and due to their dispersion profile, they can be associated with leaky modes inside the optical stack.\cite{vanlare_ACSPhotonics_2015} These kind of absorption peaks are what can be expected from using a simple PC in an weakly absorbing optical stack. 

 The bow-tie structure, $\Delta m = 1$, presents intense peaks like the chess-board, but slightly wider, and more dense, comparing Fig.~\ref{fig:absmap}(f) and (g). The four additional orders that appear in the bow-tie structure are enough to increase the bandwidth of the peaks, specially at energies close to the bandgap of GaAs (1.4 eV). However, the resonances are not a continuum, and therefore areas in the Fig.~\ref{fig:absmap}(g) with close to zero scattered absorption still remain. When increasing $\Delta m$ we can see a broadening of the peaks, increasing the absorption bandwidth, effectively creating a continuum. The penalty is a lowering of the peak intensity.

\begin{figure}
    \centering
    \includegraphics[scale=1]{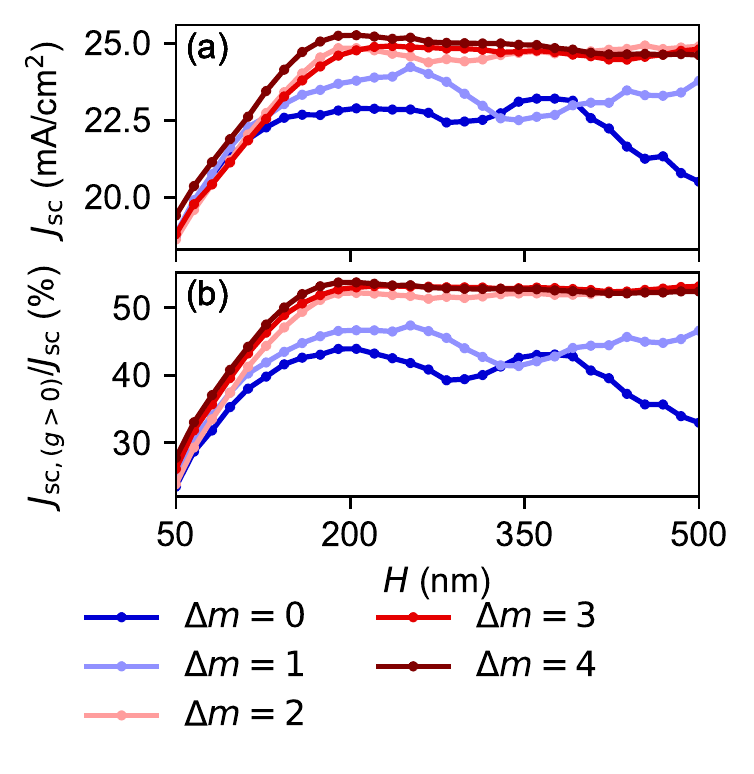}
    \caption{(a)~Generated $J_\mathrm{sc}$ of structures with $\Delta m=0,1,2,3,4$ (blue to red) for increasing PC thickness $H$. (b)~Percentage contribution to the $J_\mathrm{sc}$ from the scattered absorption $A(g>0)$ for structures with $\Delta m=0,1,2,3,4$ (blue to red).}
    \label{fig:jsc_thickness}
\end{figure}

\begin{figure}
\centering
    \includegraphics[scale=1]{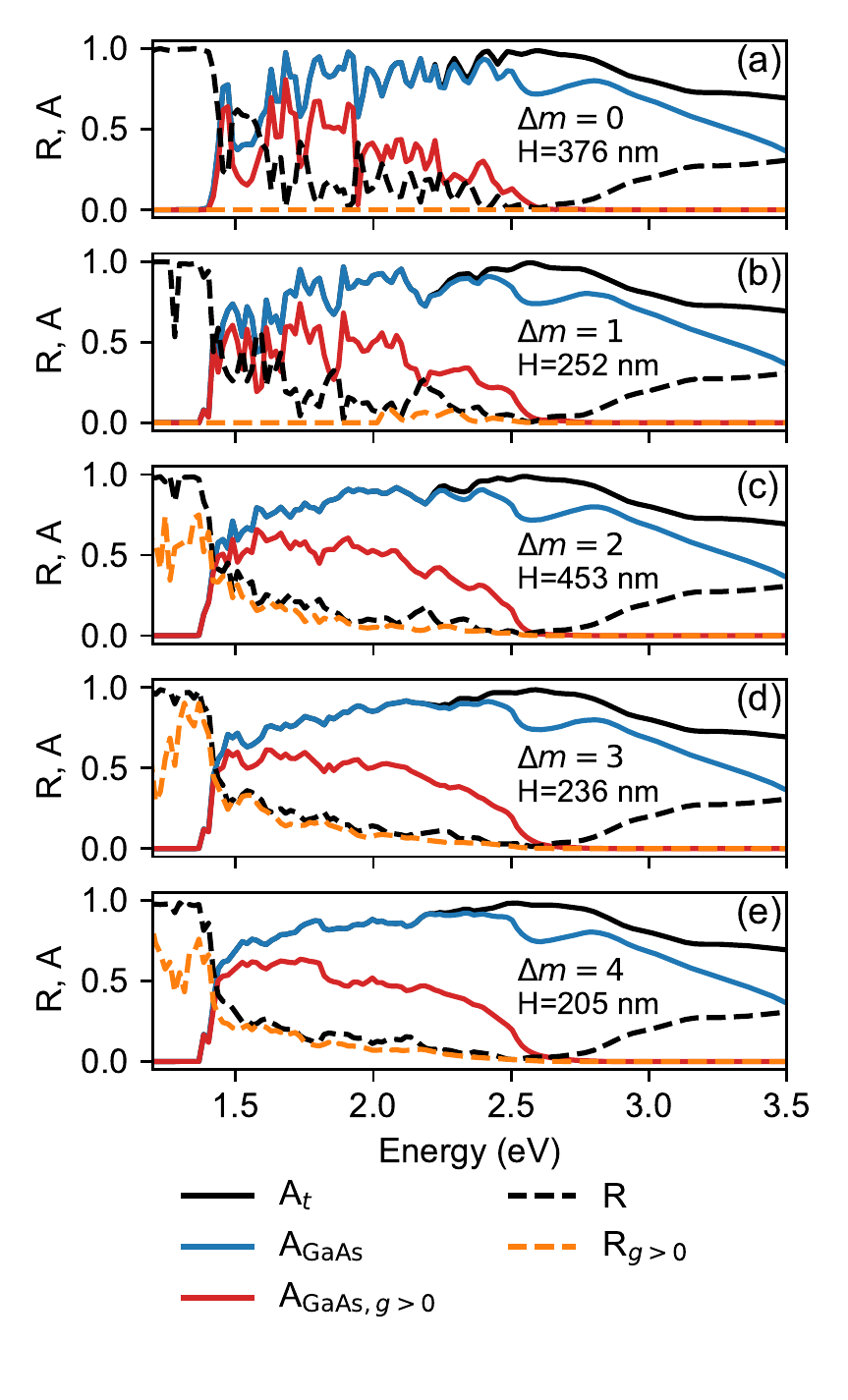}
\caption{(a)-(e) Optical response for the structures with $\Delta m=0,1,2,3,4$ respectively. Left column, (a)-(b), corresponds to the simple photonic crystals ($\Delta m=0,1$); Right column, (c)-(e) corresponds with the quasirandom structures ($\Delta m=2,3,4$). Total absorption, $A_\mathrm{t}$ (black thick), absorption in the GaAs layer ,$A_\mathrm{GaAs}$ (blue thick), scattered contribution to the absorption $A_{\mathrm{GaAs}, g>0}$ (red thick), total reflectance, $R$ (black dashed), and scattered reflectance, $R_{g>0}$ (orange dashed).}
\label{fig:abs_max}
\end{figure}

We use the absorption in Fig.~\ref{fig:absmap}(a)-(e) to calculate the generated photocurrent under the radiative limit, $J_\mathrm{sc}$, for each thickness of the PC. We use only the absorption of the GaAs layer, the absorption in other layers are considered as losses. We also study the losses in current because of the parasitic absorption outside of the GaAs active layer for energies above the bandgap (1.4 eV),\cite{buencuerpo_Opt.ExpressOE_2020}: in the silver mirror, $L_\mathrm{Ag}$, in the BSF and PC made of AlGaAs, $L_\mathrm{AlGaAs}$ and in the window layer, $L_\mathrm{w}$. However, the absorption in the inner side of the window layer may contribute on a real device, typically a 50\% of the total absorption. The maximum $J_\mathrm{sc}$ of each structure is shown in Table \ref{tab:lattice_jsc} and  the progression of the $J_\mathrm{sc}$ with $H$ is shown in Fig.~\ref{fig:jsc_thickness}. We show the total absorption ($A_\mathrm{t}$), the absorption in the GaAs layer ($A_\mathrm{GaAs}$), and the reflection spectrum ($R$) in Fig.~\ref{fig:abs_max} for the $\Delta m$ structure of maximum $J_\mathrm{sc}$. We also isolate the scattered contributions to absorption $A_{\mathrm{GaAs},g>0}$ and reflection, $R_{g>0}$.   

The five structures present an increase in the $J_\mathrm{sc}$ when increasing the thickness $H$ for the structure, see Fig. \ref{fig:jsc_thickness}. All of them present a first local maximum $H$ close to 200 nm. The increased photocurrent is due to an increased scattered contribution to the $J_\mathrm{sc}$. After the first maximum, the photocurrent decreases for the low reciprocal density structures as the square or the bow-tie ($\Delta m=0,1$), whereas it saturates for the QR structures ($\Delta m=2,3,4$), see Fig. \ref{fig:jsc_thickness}(b). The differences in the absorption are responsible of this distinction between nanostructures. The QR structures are more tolerant to thickness variations, as they fully rely on increasing the scattering, creating a continuum when the minimum thickness is reached. On the other hand, the chess-board and the bow-tie need to optimize the spectral position of the sparse resonances.

The chess-board structure ($\Delta m=0$) presents a maximum at $H= 376$ nm, not present in the other structures. This peak appears when three resonances overlap in the absorption spectrum at 1.5 eV,  Fig.~\ref{fig:absmap}(a),(f).
The absorption spectrum for this structure, see Fig.~\ref{fig:abs_max} (a) present packed, but discrete resonances associated with the resonances in $A_{\text{GaAs},(g>0)}$. However, the fine peaks, a consequence of the low PSD, limit the optimum thickness to a narrow critical value. Also, the overall contribution of the scattered absorption to the generated photocurrent is lower than for the other cases, see Fig.~\ref{fig:jsc_thickness}(b). The scattered reflection, $R_{g>0}$ is zero for the square structure, therefore is only present for the reflected ZDO. The higher reflectance is a result of the weak scattering in those wavelengths, and consequently weak absorption.

The bow-tie structure ($\Delta m =1$) increases the $J_\mathrm{sc}$ with a thicker PC until its first maximum. The $J_\mathrm{sc}$ increases with $H$ for thin structures, but the optimum height is smaller than the chess-board, $H=252 $ nm,  closer to the saturation peaks for the QR structures. The bow-tie structure presents a $J_\mathrm{sc}$ vs. $H$ in between the square structure and the QR structures. In Fig.~\ref{fig:absmap}(b),(g) we can see that the broader resonances have a maximum in the low energy region at 252 nm. 
The $R_{g=0}$ dominates the reflection as we can see when comparing  $R$ and $R_{g>0}$. This behaviour is comparable to the chess-board structure, but the $R_{g>0}$ is not zero for energies above 2.0 eV. Again, the bow-tie reflection lies in between the chess-board structure and the QR structures.

The QR structures ($\Delta m=2,3,4$) present a similar behavior, increasing $J_\mathrm{sc}$ with the thickness $H$, until it saturates.  This is expected from the absorption profiles, Fig.~\ref{fig:absmap}(c)-(e), (h)-(j) and Fig.~\ref{fig:abs_max}(c)-(e). There are small variations, around the maximum value of the $J_\mathrm{sc}$, for example $\Delta m=2$ presents a global maximum later at 453 nm. The scattered absorption is a lot more homogeneous than in the square or the bow tie for thickness above 200 nm. Once the structure diffracts enough to remove the energy from the zeroth order, there is no additional gain in increasing $H$.

The parasitic losses in the proposed design are below 1 mA/cm$^2$ for all the cases studied, see Table \ref{tab:lattice_jsc}. This is because the materials we choose are mostly transparent (10 nm AlInP window, and Al$_{0.8}$Ga$_{0.2}$As placed in the back). Also, the silver mirror is placed away from the nanostructure to avoid coupling evanescent waves to the metal. 

The end result is an increase in the overall absorption and, hence in $J_\mathrm{sc}$, see Table \ref{tab:lattice_jsc}. From these results it is possible to identify $R_{g>0}$ as the main source of losses in the QR system. Figs.~\ref{fig:abs_max} (c)-(d) show that in the range of 1.5 to 2.75 eV i) $A_\text{t}$ = $A_{\text{GaAs}}$ and $R$ = $R_{g>0}$, i.e. in this spectral range the parasitic losses and the ZDO reflectance are negligible. The $R_{g>0}$ losses appear because of the diffraction efficiency of the QR structure, even when placed on rear side of the solar cell.  Future efforts will be devoted to minimize $R_{g>0}$ to push the absorption to its ultimate limit.   

Using the IFTA for creating the photonic crystal gives us flexibility and speed for designing the objective PSD. 
In comparison, in Refs.\citenum{vanlare_ACSPhotonics_2015,
martins_NatCommun_2013} the algorithm used to solve these QR-PC structures was based on global optimizations as Monte Carlo.  Both cases converge slower than an IFTA and are more limited when using big unit cells\cite{birch_OpticsandLasersinEngineering_2000}.
The speed difference relies fundamentally in that IFTA requires very few steps to converge in comparison to global methods. For example the biggest structure studied, $\Delta m = 4$, has 199x199 pixels in the real space, (39601 variables), and it converges in only 129 steps (1.0s in a single threaded desktop computer with 4.2 GhZ). The speed to generate structures it is fundamental for machine learning approaches, where thousands of structures are needed to generate statistically significant studies.
Yu et al,\cite{yu_Sci.Rep._2017} use the Gaussian random field,
generating random waves using a Gaussian distribution in the k-space. This method is faster, but the contrast for the k-space design is lower for broadband cases\cite{yu_Sci.Rep._2017}, see supplementary materials and the demonstration code\cite{buencuerpo_GitHub_2020}. The lack of contrast limits the possibilities for using it for advanced reciprocal space engineering.

The use of QR PCs has been proposed for Si
cells~\cite{martins_NatCommun_2013,li_Sci.Rep._2015} and for thin GaAs
cells~\cite{xiao_Opt.Express_2018}. Si benefits from a thicker substrate
(microns instead of hundreds of nm) allowing the light to couple to more guided modes.
Therefore, the QR PCs obtain higher performance than the PC counterparts. In
contrast, the literature shows less benefits for the
GaAs,\cite{xiao_Opt.Express_2018} as the thickness where light trapping could be
very beneficial are few hundred nm with a few guided modes available.
In Ref.\cite{xiao_Opt.Express_2018} the silver mirror was not separated with a dielectric spacer, directly in contrast with the work presented in this paper. The proximity of the silver mirror to the ultrathin nanostructured
layer creates parasitic absorption in the metal, but it can be mitigated with a
dielectric spacer.\cite{wang_Nanolett._2012,buencuerpo_Opt.ExpressOE_2020}. The
losses using our design are negligible by looking at Fig.~\ref{fig:abs_max},
leading to high current densities for the three QR structures. Furthermore, we chose AlInP as window layer (more transparent than GaInP) and Al$_{0.8}$Ga$_{0.2}$As as PC and BSF material, again more transparent than GaInP, as presented in Ref.\cite{xiao_Opt.Express_2018}  We report here that the QR
structures are more tolerant under thickness variations than the single PCs,
with an asymptotic enhancement in the photocurrent.
\begin{figure}
    \centering
    \includegraphics[scale=1]{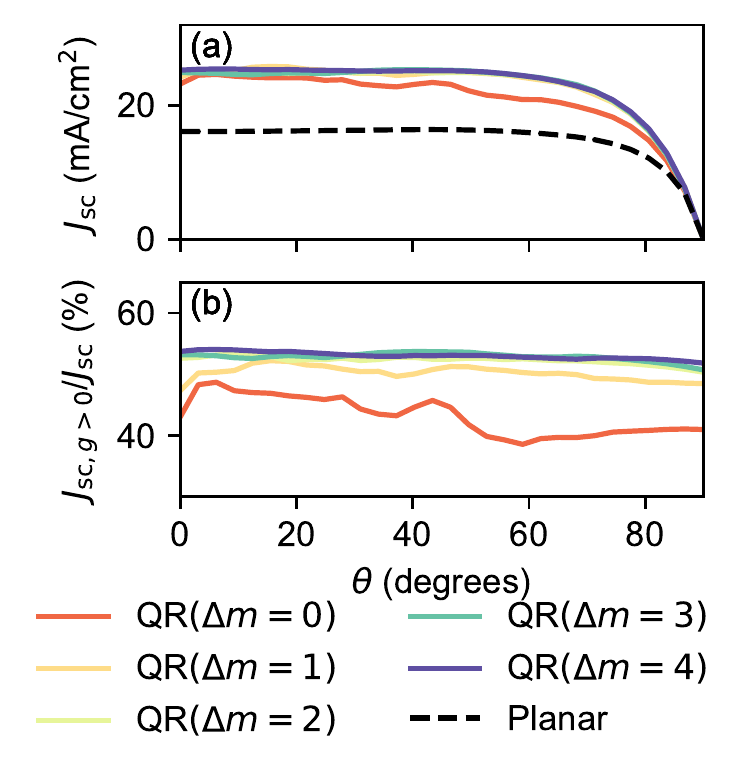}
    \caption{(a) Predicted $J_\mathrm{sc}$ of structures with $\Delta m=0,1,2,3,4$ (orange to dark blue) for increasing the incident angle from 0 to 90 degrees. The reference planar structure (black-dashed) without back photonic crystal. (b)~Percentage contribution to the $J_\mathrm{sc}$ from the scattered absorption $A(g>0)$ for structures with $\Delta m=0,1,2,3,4$ (orange to dark blue).}
    \label{fig:angle_jsc}
\end{figure}
The main results under out-of-normal incidence are shown in Fig.~\ref{fig:angle_jsc}
We study the dependence under out of normal of incidence for the five structures, see Fig.~\ref{fig:angle_jsc}.From this results one can infer the losses of our proposal to Sun position for a fix tilt mount.  In Fig.~\ref{fig:angle_jsc}(a) we observe that the chess-board structure ($\Delta m=0$) presents a lower performance under angle variations than the other structures. The simple PC relies on tuning the absorption peaks strategically to increase the absorption. The angle of incidence shifts these resonances, and therefore, the photocurrent obtained is not optimal. On the other hand, the QR structures $\Delta m=2,3,4$, present a more stable predicted photocurrent. These structures base their enhancement in the absorption in a continuum of resonances, hence being more robust than the chess-board. The bow-tie structure, again, lies in between a simple photonic crystal with an sparse reciprocal space and the QR. Analyzing the contribution of the scattered absorption to the $J_\mathrm{sc}$, Fig.~\ref{fig:angle_jsc}(b) we can observe the transition from the simple PC to the QR more clearly. The ratio between $J_{\mathrm{sc},g>0}$ and $J_\mathrm{sc}$ presents a maximum for low incident angles for the chess board structure, and then it decays as the angle of incidence increases. On the other hand, the QR structures have an almost constant percentage for all the angles studied. The QR structures are more stable in photocurrent generation, thanks to a constant enhancement in the photocurrent through the diffracted light by the structure.

In summary, the structure with the highest $J_\mathrm{sc}$ is a QR with $\Delta m=4$, yielding a $J_\mathrm{sc}$ of 25.3 mA/cm$^2$. This structure presents a more homogeneous absorption without strong resonances for all the range of spectral interest, see Fig.~\ref{fig:absmap}(a),(e) and Table \ref{tab:lattice_jsc}. The increased contribution to the absorption increases the generated photocurrent, see Table \ref{tab:lattice_jsc}, reaching values larger than 50\% for the QR-PC with $\Delta m=2,3,4$. This enhancement is robust under changes in the angle of incidence.
The $J_\mathrm{sc}$ predicted for the quasirandom structures is above the 25 mA/cm$^2$ using only 100 nm active absorber.  For comparison, the $J_\mathrm{sc}$ of an optically thick structure of 3~$\mu$m with the same ARC is 30.8 mA/cm$^2$ and a planar Ag mirror. 
Still, further optimizations are needed to achieve $J_\mathrm{sc}$ close to 30 mA/cm$^2$, like in experimental thick GaAs cells. Restricting the number of orders in the initial $Q$ matrix by engineering the reciprocal space one could obtain a compromise between the QR and the bow-tie structure, with higher and uniform $A_{\text{GaAs}, g>0}$, but with minimizing losses for $R_{g>0}$, which is the main loss.

\subsection{Statistical analysis of the generated structures}

The objective PSD is fully determined by $k_1$, $k_2$, and $\Delta m$, as explained in Section \ref{sec:design}. The fact is that many real spaces distributions exhibit a very similar PSD, in a way, the real space structure is not universally defined. \cite{wyrowski_J.Opt.Soc.Am.AJOSAA_1988} In other words, the IFTA algorithm uses a random seed when initializing the objective phase, and this seed leads to apparently different structures in the real space with similar PSD. This variation is similar to the expected variation when using direct self-assembly methods based on phase-separation.\cite{lee_PNAS_2017,zhang_Opt.ExpressOE_2018,hauser_Opt.ExpressOE_2020} It is well known that the real space geometry of the PC can modify the optical response. Therefore, there is in principle no warranty that the IFTA output would produce structures with equivalent performance, in our case the  $J_\mathrm{sc}$. We explore here the impact of this indeterminacy as the dispersion of structures could impact the performance of a device. 
We have studied the absorption and photocurrent for ten different seeds for each case $\Delta m$ from 0 to 4.
The low reciprocal density structures ($\Delta m=$ 0 and 1) do not present this variation. Therefore, we will focus the analysis on the QR structures with $\Delta m=2,3,4$.

\begin{figure*}
\begin{center}
    \includegraphics[scale=1]{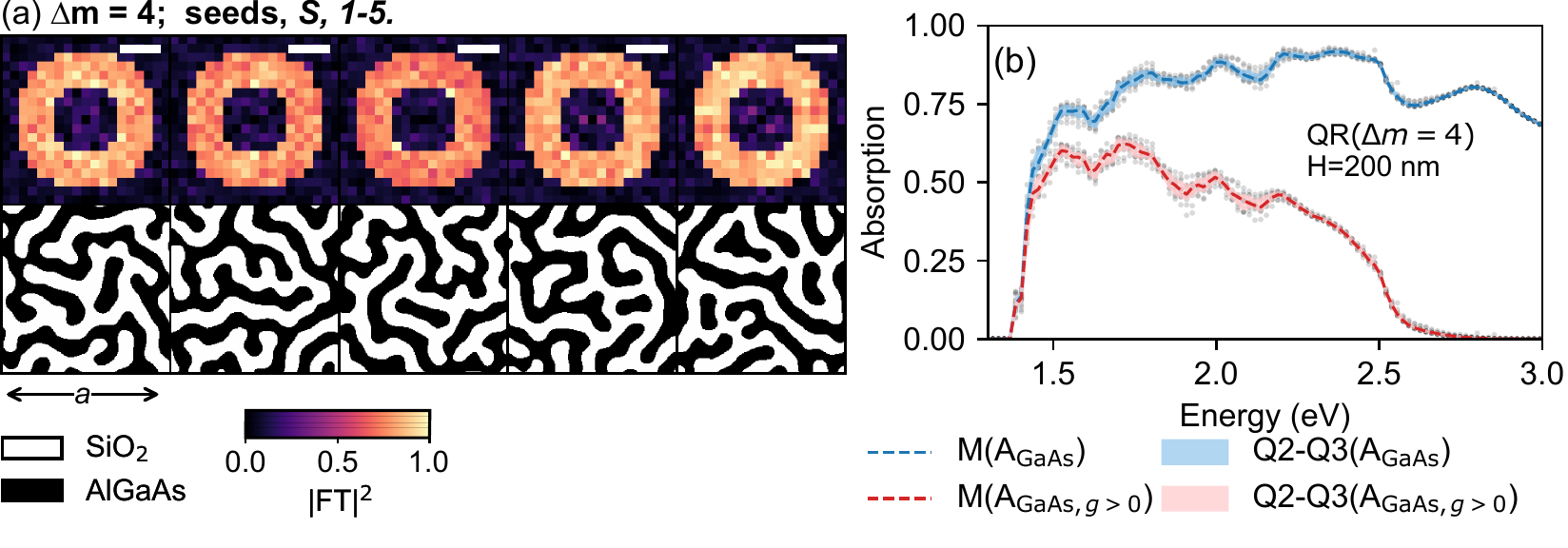}
\end{center}
\caption{(a) Reciprocal and real space for five different seeds with the same $\Delta m$ = 4. (b) Median of the distribution of absorption in the GaAs layer (blue), and scattered contribution (red) for ten different seeds for the quasirandom structures for $\Delta_m=4$. 
The absorption for each structure are shown as semitransparent gray dots. Second to third quartiles of the $A_\mathrm{GaAs}$ (pale blue filled area) and $A_{\mathrm{GaAs}, g>0}$ (pale red filled area) }
\label{fig:abs_stats}
\end{figure*}

The thickness of the structure, $H$, is kept fixed at 200 nm for all the studied structures. We chose this thickness because from the analysis of the $J_\mathrm{sc}$ we can see that for $H\geq 200$ nm the enhancement in the photocurrent saturates. We study each family of structures with equal $\Delta m$ using different seeds. We analyze the median $Med$, mean $\mu$, and standard deviation $\sigma$ of $J_\mathrm{sc}$, see Table \ref{tab:stats}. 
Also, we analyze the spectral variation of the absorption for 10 different seeds in Fig.~\ref{fig:abs_stats}(b) for $\Delta m=4$.

\begin{table}[htbp]
  \centering
  \caption{Statistic values for the predicted $J_\mathrm{sc}$ for the 10 different seeds for each QR nanostructure with increments in $\Delta_m$ using $H$=200 nm. The structures used are bitmaps in Dataset 1.} 
    \begin{tabular}{lccr}
    $\Delta m$                      & 2     & 3     & 4 \\
    \hline
    $\mu(J_\mathrm{sc})$ mA/cm$^2$    & 24.9  & 25.0  & 25.0 \\
    %$Med.(J_\mathrm{sc})$ mA/cm$^2$   & 24.9  & 24.8  & 24.9 \\
    $\sigma(J_\mathrm{sc})$ mA/cm$^2$ & 0.3   & 0.2   & 0.2 \\
    %$Max.(J_\mathrm{sc})$  mA/cm$^2$  & 25.2  & 25.4  & 25.3 \\
    %$Min.(J_\mathrm{sc})$  mA/cm$^2$  & 24.4  & 24.7  & 24.6 \\
    \hline
    \end{tabular}%
  \label{tab:stats}%
\end{table}%
Overall, we can see that the second to third quartiles (Q2-Q3) tightly envelope the median of the absorption profiles, $Med(A(E)$), see Fig.~\ref{fig:abs_stats} (b). The tightness of this envelope implies small changes in $A(E)$ when comparing between seeds, despite the changes in the unit cell of the QR. The cloud of points laying outside of the quartiles in the plot, Fig.~\ref{fig:abs_stats}(a) are more dispersed (more outliers) for the valleys in the absorption spectrum. These valleys are indeed associated with the ZDO, where the $A_{g>0}$ contribution is smaller. The ZDO is dominated by the height of the structure and the filling factor. The latter can be seen as an effective medium. But, as these structures' absorption in the long wavelength range are dominated by $A_{g>0}$, the differences in $A(E)$ are minimized and the reciprocal space defines the absorption of the structure. 

When analyzing the $J_\mathrm{sc}$, see Table \ref{tab:stats} we can see that the three cases present a mean and median almost the same as the initial seed we test, with $\mu(J_\mathrm{sc}) \approx 25$ mA/cm$^2$. The $J_\mathrm{sc}$ absolute value is defined by the reciprocal space design. Indeed, for the three cases the $\sigma(J_\mathrm{sc})$ is $\leq 0.3$ mA/cm$^2$. This deviation is comparable to real space defined structures when allowing for fabrication errors.\cite{buencuerpo_Opt.ExpressOE_2020} QR-PC structures are good candidates for self-assembly fabrication, such as spinodal decomposition based methods, because they control the reciprocal space.\cite{lee_PNAS_2017,zhang_Opt.ExpressOE_2018,hauser_Opt.ExpressOE_2020}
In summary, the thickness and reciprocal space defines the absorption of the QR structure. The real space has more degrees of freedom and therefore differences, but the final absorption is defined by the target reciprocal space: $k_1$, $k_2$ and $\Delta m$.

\section{Conclusions}
In summary, we present a method for creating photonic crystals in the reciprocal space based on the iterative Fourier transform algorithm and controlling the sparsity of the power spectral density. We apply this method to design ultrathin GaAs solar cells of 100 nm thickness with predicted photocurrent ($J_\mathrm{sc}$) of 25.3 mA/cm$^2$ for the quasirandom structure. We observe the transition from a low density reciprocal space structure, such a chess-board structure to a quasirandom structure. The additional Fourier space increases the amount of resonances, with a progression from discrete number of peaks to a continuum. For the quasirandom structures we found that the reciprocal space dominates the absorption profile and the $J_\mathrm{sc}$. Structures with different initial seeds and different real space, but similar reciprocal space, obtain comparable absorption profiles and $J_\mathrm{sc}$. 
The reciprocal space design can be modified, for example, to other types of solar cells or light-emitting diodes, by following the method described here.  The optimal spatial frequency distribution is still an open question that can be addressed in future studies taking this work as a starting point. 

\subsection*{Disclosures}
The authors declare no conflicts of interest. The data that support the findings of this study are available from Ref.\cite{buencuerpo_GitHub_2020} and from the corresponding author upon reasonable request.

\subsection*{Acknowledgments}
\footnotesize{This work was supported by the U.S. Department of Energy under Contract No. DE-AC36-08GO28308 with Alliance for Sustainable Energy, LLC, the Manager and Operator of the National Renewable Energy Laboratory. Funding provided by U.S. Department of Energy Efficiency and Renewable Energy Solar Energy Technologies Office under Agreement Numbers 34911 and 34358. This research was performed using computational resources sponsored by the Department of Energy's Office of Energy Efficiency and Renewable Energy and located at the National Renewable Energy Laboratory. The U.S. Government retains and the publisher, by accepting the article for publication, acknowledges that the U.S. Government retains a nonexclusive, paid up, irrevocable, worldwide license to publish or reproduce the published form of this work, or allow others to do so, for U.S. Government purposes. JML acknowledges the financial support of Spanish MINECO (RYC-2017-21995) and the European Union FSE. JMR acknowledges support by MCINN (ENE2017‐91092‐EXP, RTI2018-096937-B-C22) and Com. Mad. (P2018/EMT-4308).}
\bibliography{QR_2020}
\end{document}